\documentclass[%
aps,
pre,
preprint,%
superscriptaddress,
floatfix
]{revtex4-1}
\usepackage{graphicx}
\usepackage{dcolumn}
\usepackage{bm}
\usepackage[utf8]{inputenc}
\usepackage{epstopdf}
\usepackage{amsmath}
\usepackage{amsthm}
\usepackage{float}
\usepackage{microtype}
\usepackage{color}

\usepackage{hyperref}
\usepackage[capitalize]{cleveref}
\usepackage{wrapfig}
\usepackage{float}
\usepackage{color}

\def\be{\begin{equation}}
\def\ee{\end{equation}}
\def\beq{\begin{eqnarray}}
\def\eeq{\end{eqnarray}}
\def\l{\left}
\def\r{\right}

\makeatother

\begin{document}

\title{Plasma Turbulence in the Scrape-off Layer of the ISTTOK Tokamak}

\author{Rogério Jorge}\email{rogerio.jorge@epfl.ch}
\affiliation{École Polytechnique Fédérale de Lausanne (EPFL), Swiss Plasma Center (SPC), CH-1015
Lausanne, Switzerland}
\affiliation{Instituto de Plasmas e Fusão Nuclear, Instituto Superior Técnico, Universidade de Lisboa, 1049-001 Lisboa, Portugal}
\author{Paolo Ricci}
\affiliation{École Polytechnique Fédérale de Lausanne (EPFL), Swiss Plasma Center (SPC), CH-1015
Lausanne, Switzerland}
\author{Federico D. Halpern}
\affiliation{École Polytechnique Fédérale de Lausanne (EPFL), Swiss Plasma Center (SPC), CH-1015
Lausanne, Switzerland}
\author{Nuno F. Loureiro}
\affiliation{Plasma Science and Fusion Center, Massachusetts Institute of Technology, Cambridge, Massachusetts 02139, USA}
\author{Carlos Silva}
\affiliation{Instituto de Plasmas e Fusão Nuclear, Instituto Superior Técnico, Universidade de Lisboa, 1049-001 Lisboa, Portugal}

\begin{abstract}
The properties of plasma turbulence in a poloidally limited scrape-off layer (SOL) are addressed, with focus on ISTTOK, a large aspect ratio tokamak with a circular cross section. Theoretical investigations based on the drift-reduced Braginskii equations are carried out through linear calculations and non-linear simulations, in two- and three-dimensional geometries. The linear instabilities driving turbulence and the mechanisms that set the amplitude of turbulence as well as the SOL width are identified. A clear asymmetry is shown to exist between the low-field and the high-field sides of the machine. While the comparison between experimental measurements and simulation results shows good agreement in the far SOL, large intermittent events in the near SOL, detected in the experiments, are not captured by the simulations.
\end{abstract}

\maketitle

\section{Introduction}
\label{intro_section}

In recent years, significant progress was made in the study of the plasma turbulence properties in the scrape-off layer (SOL) of tokamaks \cite{Ricci2015}, the region that exhausts the tokamak power, controls the plasma fueling and the impurity dynamics, and plays  a major role in determining the overall plasma confinement \cite{Eich2013, Kukushkin2011,Divertor2002,Divertor2002,Lipschultz2007}.
These theoretical investigations \cite{DIppolito2011} focused mainly on the toroidally limited SOL \cite{Ribeiro2008,Tamain2014,Ricci2013a}, a configuration that is relevant to the ITER start-up and ramp-down phases during which the inner or the outer vessel wall will be used as the limiting surface \cite{Arnoux2013,Jackson2009}.
In this scenario, using low-frequency fluid models, the turbulent regimes were identified. It was found that drift waves (DW) and ballooning modes (BM) drive the plasma turbulent dynamics, with the resistive BM being the main drive in typical existing tokamak conditions \cite{Mosetto2012}, a result in agreement with previous experimental results \cite{LaBombard2005,Labombard2008}. Simulations and analytical estimates revealed that the fluctuations saturate due to a local flattening of the plasma gradients and associated removal of the linear instability drive \cite{Ricci2013a}. By using a balance between turbulent transport and parallel losses at the vessel, a scaling of the pressure scale length was derived. A thorough comparison with experimental measurements was carried out with significant success \cite{Halpern2015}. The question of how these findings can be applied to other configurations remains open and is one of the main motivations of this work.

The goal of the present paper is the study of turbulence properties in a poloidally limited geometry, such as the one of ISTTOK \cite{Varandas1996,Silva2009}, a large aspect ratio tokamak ($R/a \sim 5.4$, where $R$ and $a$ are the major and minor radius respectively) with a circular cross section. By intercepting the magnetic field lines on a poloidal plane, a poloidal limiter avoids the connection between the low- and the high-field sides of the machine. This allows the turbulent properties, and therefore the pressure scale length and the SOL width, to retain a strong poloidal dependence. The shorter connection length, with respect to the toroidally limited case, leads to enhanced parallel losses, steepening the gradients and, as we show, changing the relative role of DW and BM in driving turbulence.

We carry out our investigation by using linear and non-linear simulations, in two- and three-dimensional geometries, that are based on the drift-reduced Braginskii equations \cite{Zeiler1997}. These are solved with GBS \cite{Ricci2012,Halpern2016a}, a numerical simulation code developed with the goal of simulating plasma SOL turbulence by evolving the full profiles of the various plasma quantities with no separation between perturbations and equilibrium, and was validated against experiments such as the TORPEX device \cite{Ricci2009} and several other machines \cite{Halpern2015}, verified with the method of manufactured solutions \cite{Riva2016}, and benchmarked against other major SOL simulation codes, including BOUT++ \cite{Dudson2009}, HESEL \cite{Nielsen2015}, and TOKAM3X \cite{Tamain2014}. The parameters of our study rely on the ones from ISTTOK, where a clear asymmetry between the low and the high field sides was found \cite{Silva2011a}. We uncover the instabilities driving turbulence and the turbulent regimes in ISTTOK, and we quantitatively compare our simulation and theoretical results with some of the measurements taken in this device.

This paper is organized as follows. Section \ref{model_section} describes the model equations and the ISTTOK simulation results. In Sec. \ref{sec:identification} we investigate the nature of the instabilities driving turbulence in a poloidally limited SOL. Sec. \ref{sat_section} discusses the development of the linear instabilities into non-linear turbulence and provides an estimate of the time-averaged pressure gradient scale length. Finally, in Sec. \ref{exp_comparison}, a comparison between ISTTOK experimental measurements and simulations is reported. The conclusions are presented in Sec. \ref{sec:conclusion}.

\section{Model Equations and ISTTOK Simulation Results}
\label{model_section}

In the ISTTOK SOL, the turbulent time scales (such as the one measured by Langmuir probes $\lesssim 10^{-5}$ s) are slower than the collisional time ($\tau_e \sim 10^{-6}$ s), and the scale lengths along the (poloidally limited) magnetic field ($L_\parallel = 2 \pi R \sim 3$ m) are longer than the mean free path ($\lambda_{\text{mfp}} \sim 1$ m). This implies that the plasma distribution function is close to a local Maxwellian \cite{Braginskii1965}, and justifies the use of a fluid description. Furthermore, the turbulent time scales are slower than the ion cyclotron time ($\omega_{ci}^{-1} \sim 10^{-7}$ s), and the perpendicular scale lengths ($L_p \sim 1$ cm) are longer than the ion gyroradius ($\rho_i \sim 0.1$ cm). It follows that a description of the ISTTOK SOL based on the three-dimensional, two-fluid, drift-reduced Braginskii equations can be used \cite{Zeiler1997}. According to Ref. \cite{LaBombard2005}, electromagnetic effects lead to a non-negligible enhancement on heat and particle transport in the SOL. At the value of the MHD ballooning parameter $\alpha_{\text{MHD}} = \beta_e R/L_p \sim 1.2 \times 10^{-3}$ in ISTTOK, we do not expect the ideal ballooning mode to play a major role. We refer the reader to Ref. \cite{Halpern2013a} for a detailed treatment of electromagnetic effects in the SOL within the drift-reduced fluid description and here we consider the electrostatic limit. The model equations are

\begin{align}
 \frac{\partial n}{\partial t} =& -\frac{c}{B}[\phi,n]+\frac{2c}{eB}\Big[C(n T_e)-enC(\phi)\Big]
 -\nabla_{\parallel}\left(n V_{\parallel e}\right)+\mathcal{D}_n(n)+S_n,\label{eq:cont_ph}\\
\frac{\partial \Omega}{\partial t}=& -\frac{c}{B}[\phi,\Omega]-V_{\parallel i}\nabla_{\parallel}\Omega+\frac{\omega_{ci} }{3en}C(G_i)
+\mathcal{D}_\omega(\omega)\nonumber\\
&+\frac{m_i \omega_{ci}^2}{e n}\l[\nabla_{\parallel}\l(n(V_{\parallel i}-V_{\parallel e})\r)
+\frac{2}{m_i \omega_{ci}}C(n (T_i+T_e))\r],\label{eq:vort_ph}\\ 
 m_e \frac{\partial V_{\parallel e}}{\partial t}=& -m_e\frac{c}{B}[\phi,V_{\parallel e}]-m_e V_{\parallel e}\nabla_{\parallel}V_{\parallel e}-1.71{\nabla_{\parallel}T_e}\nonumber\\
 &-\frac{2}{3 n}\nabla_{\parallel}G_e-0.51 m_e \nu_e(V_{\parallel e}-V_{\parallel i})
 +e\nabla_{\parallel} \phi-{T_e}{\nabla_{\parallel}\ln n}+{\mathcal{D}_{V_{\parallel e}}(V_{\parallel e})/n},\label{eq:ohm_ph}\\
 m_i \frac{\partial V_{\parallel i}}{\partial t} =& -m_i \frac{c}{B}[\phi,V_{\parallel i}]-m_i V_{\parallel
i}\nabla_{\parallel}V_{\parallel i}-\frac{2}{3n}{\nabla_{\parallel}G_i}
-{\nabla_{\parallel}\left[n (T_e+T_i)\right]}/n+{\mathcal{D}_{V_{\parallel i}}(V_{\parallel
i})}/n,\\
 \frac{\partial T_e}{\partial t} =& -\frac{c}{B}[\phi,T_e]+\frac{4cT_e}{3eB}\left[\frac{7}{2}C(T_e)+\frac{T_e}{n}C(n)
-eC(\phi)\right]+0.71\frac{2T_e}{3}\l[(V_{\parallel i}-V_{\parallel e})
{\nabla_{\parallel}\ln n}\r.\nonumber\\
&\l.+\nabla_{\parallel}\left(V_{\parallel i}-2.4 V_{\parallel
e}\right)\r]-V_{\parallel
e}\nabla_{\parallel}T_e
+\mathcal{D}_{T_e}(T_e)+S_{T_e}\label{eq:te_ph},\\
\frac{\partial T_i}{\partial t}=&-\frac{c}{B}\left[\phi,T_i\right]+\frac{4c}{3eB}\frac{T_i}{n}\left[C(n T_e)-e n C(\phi)\right]
+ \frac{2}{3}{T_i}\left(V_{\parallel i}-V_{\parallel e}\right)\nabla_{\parallel} \ln n\nonumber\\
&-\frac{2}{3}T_i\nabla_{\parallel}V_{\parallel e}-V_{\parallel i} \nabla_{\parallel}T_i
-\frac{10cT_i}{3e B}C(T_i)+\mathcal{D}_{T_i}(T_i)+S_{T_i}.\label{eq:ti_ph}
\end{align}

\noindent where $\Omega = \omega + \nabla^2_{\bot}T_i/e$, with $\omega=\nabla_\perp^2 \phi$ the vorticity and $\phi$ the electrostatic potential. In the density ($n$) and electron and ion temperature ($T_e$, $T_i$) equations, source terms $S_{n,T}=S_{0n,T}\exp\l[-{(x-x_s)^2}/{\sigma_s^2}\r]$ are added to mimic the plasma outflow from the core into the SOL.
The diffusion operators for a generic field $A$, defined as $\mathcal{D}_A(A) = \chi_A \nabla^2_\perp A$, are present for numerical reasons, i.e., to damp fluctuations at the grid scale.
The gyroviscous terms $G_{i,e}$ are defined as

\begin{equation}
    G_{i,e}=-\eta_{0 i,e}\l\{{2}\nabla_\parallel V_{\parallel i,e} +\frac{c }{e n B}\l[{e n C(\phi)}\pm C(n T_{i,e})\r]\r\},
\end{equation}

\noindent with $\eta_{0 i,e}$ the Braginskii's viscosity coefficients \cite{Braginskii1965}. In Eqs. (\ref{eq:cont_ph} - \ref{eq:ti_ph}), we have also introduced the magnetic field unit vector $\bm b = \bm B/B$, the curvature operator $C(f) =({B}/{2})\nabla \times \l({\bm b}/{B}\r) \cdot \nabla f$, and the Poisson brackets operator $[\phi, f] =\bm b \cdot (\nabla \phi \times \nabla f)$. We use the Spitzer's estimate of the electron-ion collision frequency, that is $\nu_e=2.91 \times 10^{-6} \lambda n T_e^{-3/2}$, with $\lambda$ the Coulomb logarithm, $T_e$ in eV, and $n$ in cm$^{-3}$.

For simplicity, we consider a large aspect ratio geometry, and no magnetic shear. An orthogonal coordinate system $[y, x, z]$ is used, where $x$ is the flux coordinate corresponding to the radial direction, $z$ is a coordinate along the magnetic field $\bm{B}$, and $y$ is the coordinate perpendicular to both $x$ and $z$. Because of the considered large aspect ratio limit, the plane $(x, y)$ coincides with the poloidal plane, which implies $y = a\theta$, where $\theta$ is the poloidal angle ($-\pi < \theta < \pi$), with $\theta=0$ corresponding to the low-field side (LFS) equatorial midplane and $\theta=\pm \pi$ to the high-field side (HFS). In the rest of the paper, we use $\theta$ and $\varphi$ as the poloidal and toroidal coordinates respectively, with $z = R \varphi / \cos \epsilon$, where $\epsilon$ is the magnetic field pitch angle $\epsilon = \arctan(a/qR)$ and $q$ the safety factor. The parallel gradient is $\nabla_\parallel = \partial_z \simeq R^{-1}(\partial_\phi + q^{-1} \partial_\theta)$, and the perpendicular Laplacian is $\nabla^2_\perp = \partial^2_x+a^{-2}\partial^2_\theta$. The poloidal limiter is located at $\varphi = 0, 2\pi$, where we impose the Bohm sheath conditions for the ion and electron parallel velocities as $V_{\parallel i}=\pm c_s$ and $V_{\parallel e}  = \pm c_s \exp({\Lambda-{e \phi}/{T_e}})$ respectively, with $c_s = \sqrt{(T_e+T_i)/{m_i}} $ and $\Lambda=0.5 \ln\l[ {{m_i}/(2\pi{m_e})}\r] \simeq 3$ \cite{Loizu2012a}.

To solve Eqs. (\ref{eq:cont_ph} - \ref{eq:ti_ph}) we use GBS, a code that was developed in the past few years to simulate the turbulent dynamics in the tokamak SOL \cite{Ricci2012,Halpern2016a}. We perform a simulation (denoted as the standard ISTTOK simulation in the following) whose parameters follow the ones of the ISTTOK tokamak, which has a major radius $R = 0.46$ m, minor radius $a = 0.085$ m, and a toroidal magnetic field $B_T = 0.5$ T. We express the input parameters and the simulation results in terms of the ISTTOK's last closed flux surface parameters, i.e., a reference electron temperature $T_{e0} =$ 20 eV, density $n_0=10^{18}$ m$^{-3}$, magnetic field $B=0.5$ T, and ion sound Larmor radius $\rho_{s0} \equiv c_{s0}/\omega_{ci} \simeq 0.9$ mm [where $c_{s0}=\sqrt{T_{e0}/m_i}$ and $\omega_{ci} = e B/(m_i c)$]. This results in $R \simeq 504$ $\rho_{s0}$, $a \simeq 93$ $\rho_{s0}$, dimensionless resistivity $\nu  = {e^2 n_0 R}/({m_i \sigma_{\parallel 0} c_{s0}}) \simeq 1 \times 10^{-3}$ [where $\sigma_\parallel = 1.96~n e^2 / (m_e \nu_e)$ is the parallel conductivity], mass ratio $m_i/m_e \simeq 5\times 10^{-4}$, and safety factor $q \simeq 8$.
As there are no detailed measures of the ion temperature, we perform our non-linear simulations in the cold ion limit ($\tau = T_i/T_e=0$), and analyze the effect of finite $T_i$ on the linear growth rate of the unstable modes and the time-averaged pressure gradient length in \cref{sat_section}.

The simulation has a radial extension $0 < x < 50$ $\rho_{s0}$. The plasma and heat sources, located at $x_s = 10$ $\rho_{s0}$, have a characteristic width of $\sigma_s = 2.5$ $\rho_{s0}$. Our analysis considers only the physically meaningful region $x > x_s$. We remark that ISTTOK's radial distance between the last closed flux surface and the outer wall is approximately $16$ $\rho_{s0}$, in practice comparable to the experimental SOL width. Since a set of boundary conditions that properly describes the interaction of the plasma with the outer wall is not known, we consider a radial domain extension larger than in the experiment, so that the plasma pressure decays to a negligible value at the outer wall, and the boundary conditions we impose at this location have a negligible impact on the turbulent properties. Specifically, at $x=0$ and $x=50$ $\rho_{s0}$, Neumann boundary conditions are used for density, temperature, electric potential, while Dirichlet boundary conditions are used for the vorticity.
By computing the power spectrum of the fluctuations, we observe that $\chi_A\ge 6c_{s0}/(\rho_{s0}^2 R)$ properly damps fluctuations at the grid scale. Moreover, the simulation results are not sensitive to the values of the diffusion coefficients for the range of values $6 < \chi_A \rho_{s0}^2 R/c_{s0} < 20$, so the value of $\chi_A=12 c_{s0}/(\rho_{s0}^2 R)$ is used for all fields. A spatial grid of $512 \times 64 \times 32$ and a time step of $10^{-4} R/c_{s0}$ is employed.

\begin{figure}
	\centering
	\includegraphics[width=.7\linewidth]{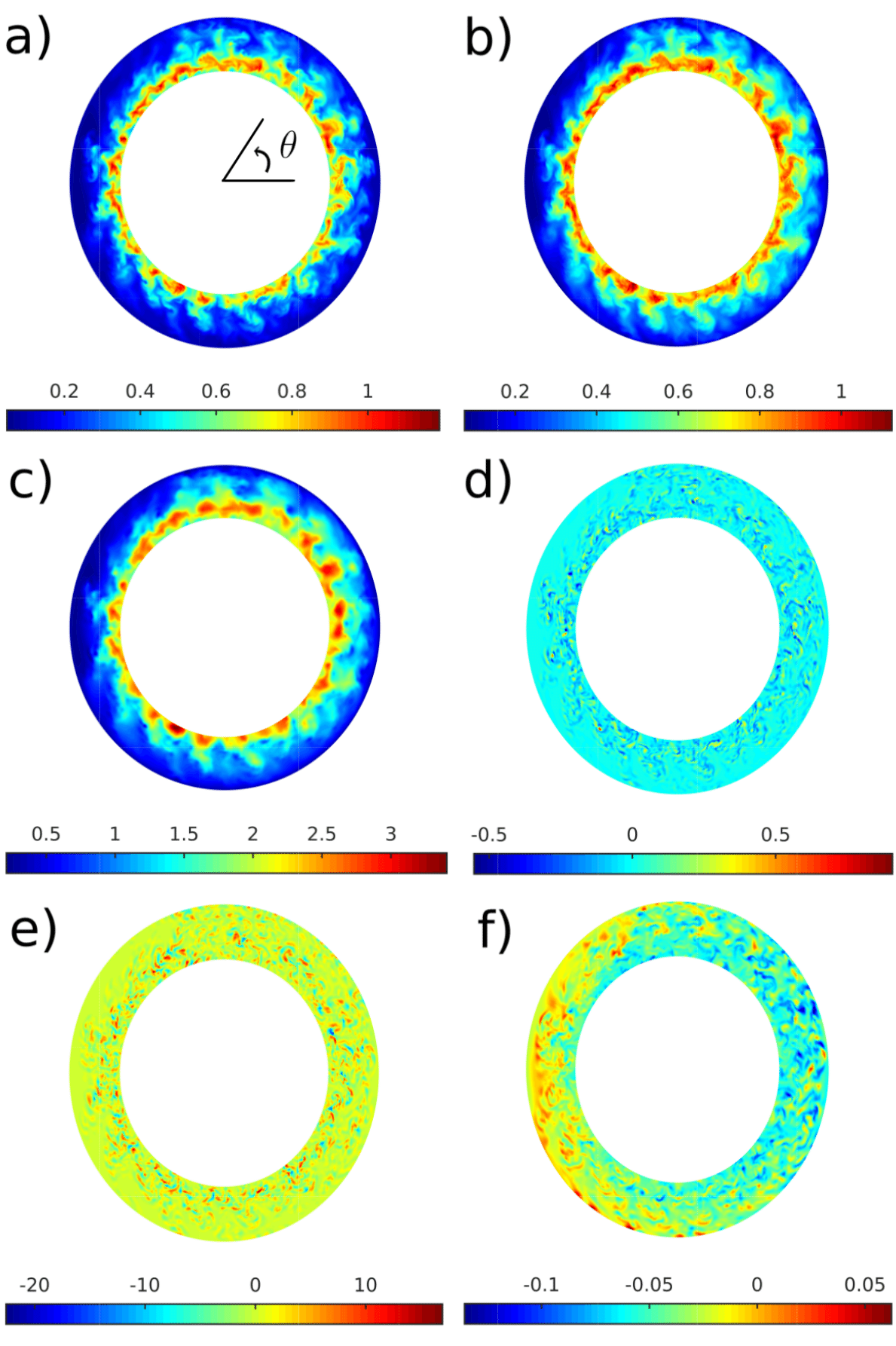}
	\caption{Snapshots of plasma turbulence in the standard ISTTOK simulation on a poloidal cross section halfway between the limiter plate ($\varphi = \pi$). We show: (a) plasma density $n/n_0$, (b) electron temperature $T_e/T_{e0}$, (c) electrostatic potential $\phi/e {T_{e0}}$, (d) vorticity $\omega=\rho_{s0}^2 \nabla^2_\perp \phi/e {T_{e0}}$, (e) electron $V_{\parallel e}/c_{s0}$, and (f) ion $V_{\parallel i}/c_{s0}$ parallel velocities.}
	\label{poloidalgbs}
\end{figure}

\begin{figure}
	\centering
	\includegraphics[width=.7\linewidth]{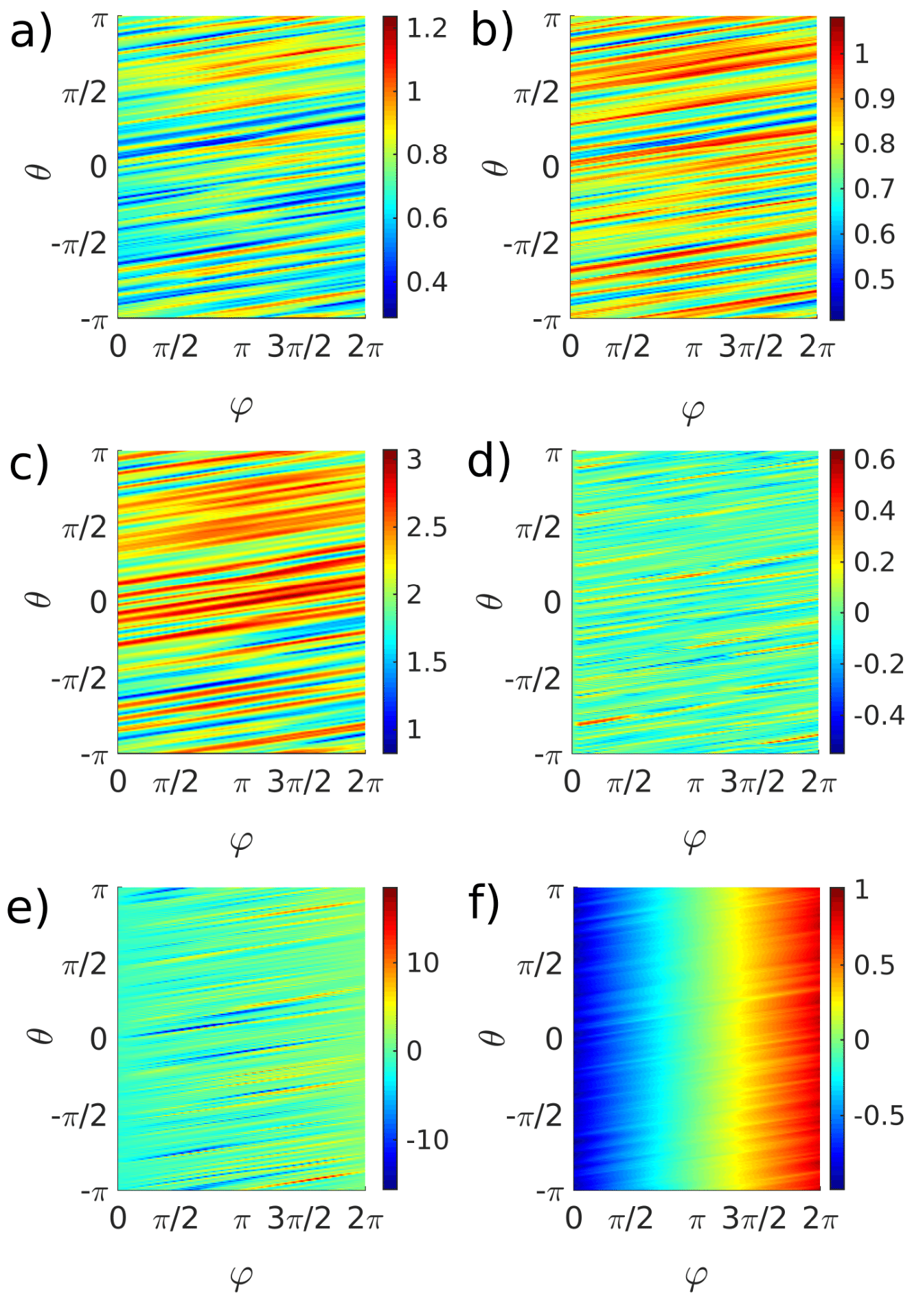}
	\caption{Snapshots of plasma turbulence for the standard ISTTOK simulation on a toroidal cross section at $x = x_s + 5$ $\rho_{s0}$. We show: (a) plasma density $n/n_0$, (b) electron temperature $T_e/T_{e0}$, (c) electrostatic potential $\phi/e {T_{e0}}$, (d) vorticity $\omega=\rho_{s0}^2 \nabla^2_\perp \phi/e {T_{e0}}$, (e) electron $V_{\parallel e}/c_{s0}$, and (f) ion $V_{\parallel i}/c_{s0}$ parallel velocities.}
	\label{phithetagbs}
\end{figure}

A typical turbulent snapshot for the standard ISTTOK simulation is shown in Figs. \ref{poloidalgbs} and \ref{phithetagbs}.
Figure \ref{poloidalgbs} shows the development of the plasma turbulence on the poloidal plane $\varphi = \pi$ midway between the two sides of the limiter plate. We observe that $n$, $T_e$ and $\phi$ fluctuations are stronger on the LFS, $\theta=0$, compared to the HFS, $\theta=\pm\pi$, where the SOL width is narrower. Figure \ref{phithetagbs}, taken at a toroidal plane $x=x_s + 5$ $\rho_{s0}$, confirms that turbulent fluctuations tend to be aligned to the magnetic field lines. The ion parallel velocities $V_{\parallel i}$ are $-c_s$ and $+c_s$ at the limiter plates $\varphi=0$ and $2\pi$ respectively, and the $V_{\parallel e}$ fluctuations are much larger due to the small electron inertia.

\section{Identification of Driving Linear Instabilities}
\label{sec:identification}

Previous studies on the drift-reduced Braginskii equations show that ballooning modes (BM) and drift waves (DW) are the instabilities that drive most of the transport in a toroidally limited SOL \cite{Hastie2003, Mosetto2012}. 
BM are driven unstable by magnetic field line curvature and plasma pressure gradients. They are characterized by a large ($\sim \pi/2$) phase shift between $n$ and $\phi$ \cite{Shchepetov2013}, and their growth rate is maximum at the longest parallel wavelength allowed in the system. On the contrary, DW arise at finite $k_\parallel$ due to the $\bm E \times \bm B$ convection of the pressure profile, and are driven unstable by finite resistivity and electron inertia, showing an adiabatic electron response, and a small phase shift between $n$ and $\phi$ \cite{Mosetto2013}.
Besides BM and DW, the Kelvin-Helmholtz (KH) instability, driven by shear flows, and the sheath mode, driven by a temperature gradient when magnetic field lines terminate on a solid wall and sheath physics plays a role, may also influence the SOL dynamics \cite{Rogers2010}.

\begin{figure}
	\centering
	\includegraphics[width=.99\linewidth]{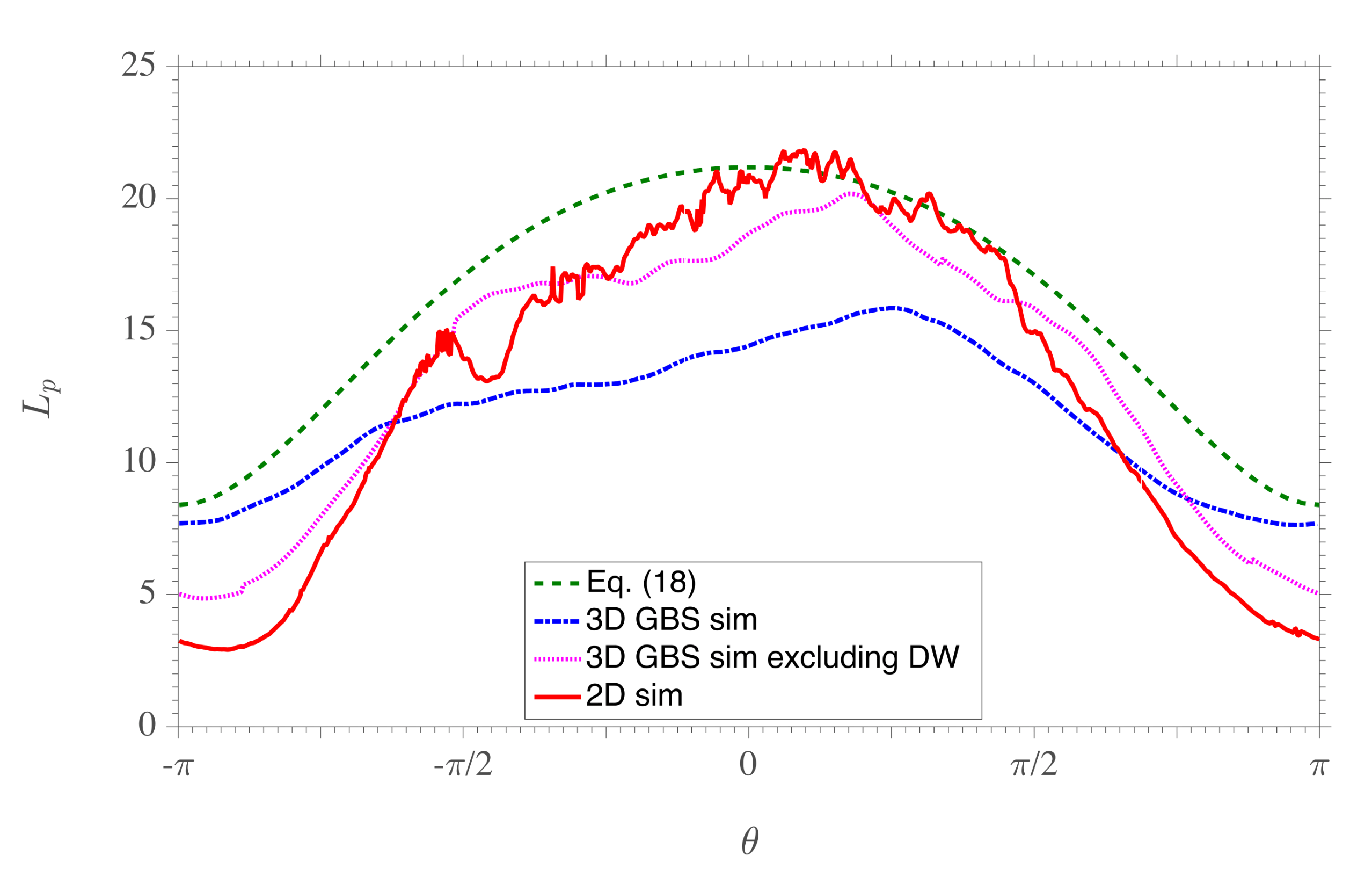}
	\caption{The equilibrium pressure scale length, $L_p$, is plotted as a function of the poloidal angle, $\theta$, from an exponential fit in the radial direction of the type $p/p_0=e^{-x/L_p(\theta)}$ of the two- (red) and three-dimensional simulations, with (blue) and without (purple) DW, and the prediction from \cref{gr10} (green).}
	\label{lpfig}
\end{figure}

The role of DW in the system is assessed by two different studies. First, we compare the standard ISTTOK simulation with a two-dimensional simulation carried out with a model that, having excluded $k_\parallel \not= 0$ modes (and in particular DW), evolves the field-line averaged density, $n(r,\theta)$, potential, $\phi(r,\theta)$, and temperature, $T(r,\theta)$ (see Ref. \cite{Ricci2011}).
Second, we perform a three-dimensional simulation where we exclude DW dynamics by neglecting the diamagnetic terms, $T_e \nabla_\parallel \ln n$ and $1.71 \nabla_{\parallel} T_e$, in Ohm's law, \cref{eq:ohm_ph}. The results of these experiments cast in terms of the averaged pressure gradient scale length $L_p \simeq |p/\nabla p|$ (where $p=n T_e$), are compared in \cref{lpfig} with the result from the full 3D GBS simulations. This includes the standard ISTTOK simulation (blue line), and the two- and three-dimensional simulations that exclude the DW dynamics (red and purple lines, respectively).
Motivated by the difference between the LFS and the HFS following the removal of DW in \cref{lpfig}, we analyse separately the different poloidal positions. We note that this is justified by the fact that the plasma rotates poloidally on a time scale ${2 \pi a}/{V_{E \times B}} \sim 2 \pi a L_p \omega_{ci}/ (\Lambda c_{s0}^2) \sim 10^{-3}$ s, which is much slower than the turbulent time scales ($\sim 10^{-5}$ s).

We start our analysis at the LFS. Here, curvature is unfavourable, BM are expected to be unstable and, comparing the standard GBS simulation with the one excluding DW in \cref{lpfig}, it is observed that removing DW from the system leads to increasing values of $L_p$, suggesting that these may have a significant role.
By linearizing the drift-reduced Braginskii system of equations (\ref{eq:cont_ph} - \ref{eq:ti_ph}) in the cold-ion limit, assuming background density and temperature profiles with radial scale lengths given by $L_n$ and $L_{T_e}$ respectively, and a perturbation of the form $e^{\gamma t + i k_y y+i k_\parallel z}$, we obtain the following dispersion relation that captures DW and BM

\begin{align}
\begin{split}
        g {\frac{\gamma^2}{\omega_{ci}^2}}\frac{k_y^2}{k_\parallel^2} \frac{m_e}{m_i}&=i\frac{k_y\rho_{s0}^2}{L_n}\frac{\omega_{ci}}{\gamma}(1+1.71 \eta_e)-2.95 g k_y^2 \rho_{s0}^2 -1,
\end{split}
\label{drbmdw}
\end{align}
\noindent with
\begin{align}
\begin{split}
        g&=\frac{1-{2 \cos \theta}(1+\eta_e)(\rho_{s0}^2/R L_n)(\omega_{ci}^2/\gamma^2)}{1+{4.28 i}\cos \theta (k_y \rho_{s0}^2/R)(\omega_{ci}/{\gamma})},
\end{split}
\end{align}

\begin{figure}
	\centering
	\includegraphics[width=.99\linewidth]{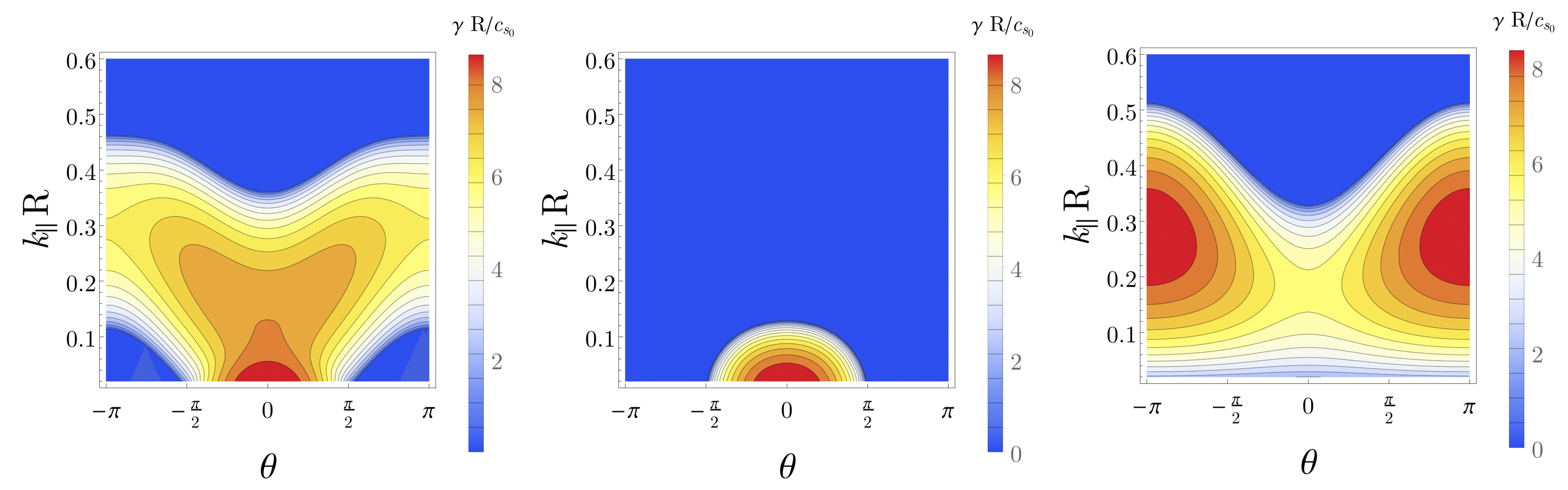}
	\caption{Linear growth rate as a function of the poloidal angle, $\theta$, and of the parallel wavenumber normalized to the major radius $k_\parallel R$. From left to right: the solution of the full dispersion relation that couples inertial DW and BM, \cref{drbmdw}; the solution of \cref{drbm} for the pure BM; and the solution of \cref{drdw} for the pure DW.}
	\label{gammakytheta}
\end{figure}

\noindent and $\eta_e = L_n/L_{T_e}$. We remark that, to deduce \cref{drbmdw}, we also take into account the fact that $\theta$ is almost constant along a field line due to a high $q$ at the edge, we neglect both sound wave coupling and compressibility terms in the continuity (\ref{eq:cont_ph}) and temperature (\ref{eq:te_ph}) equations, since $\gamma \gg k_\parallel c_s$ and $L_\parallel/R \ll 1$ (as confirmed by the linear analysis below), and we focus on the inertial limit by neglecting the resistivity term $\nu_e$ in Ohm's law (\ref{eq:ohm_ph}). The inertial nature of the instabilities present in the system is confirmed in Sec. \ref{sat_section}.

The largest growth rate solution of \cref{drbmdw} is plotted as a function of $k_\parallel$ and $\theta$ in the left panel of Fig. \ref{gammakytheta}, having chosen $L_n$, $\eta_e$,  and $k_y \rho_{s0}$ according to the results of the ISTTOK standard simulation. This growth rate is compared with the maximum one resulting from the dispersion relation of the pure BM,

\begin{equation}
\begin{split}
        \frac{\gamma^2}{\omega_{ci}^2}-{2 \cos \theta}(1+\eta_e)\frac{\rho_{s0}^2}{R L_n}=-\frac{k_\parallel^2}{k_y^2} \frac{m_i}{m_e},
        \label{drbm}
\end{split}
\end{equation}
and pure DW,
\begin{equation}
\begin{split}
        {\frac{\gamma^2}{\omega_{ci}^2}}\frac{k_y^2}{k_\parallel^2} \frac{m_e}{m_i}&=i\frac{k_y\rho_{s0}^2}{L_n}\frac{\omega_{ci}}{\gamma}(1+1.71 \eta_e)-2.95 k_y^2 \rho_{s0}^2 -1.
        \label{drdw}
\end{split}
\end{equation}

One observes from \cref{gammakytheta} that pure BM are unstable for $k_\parallel R < 0.15$ and for $k_\parallel=0$ they exhibit a strong growth rate at the LFS. However, as they are strongly stabilised by finite $k_\parallel$, at the typical values of $k_\parallel \sim$ 0.1 - 0.2 found in the standard ISTTOK simulations, DW are the fastest growing instability. We note that the enhancement of transport observed in \cref{lpfig} when DW are removed is due to the increased size of the turbulent eddies.

To conclude the analysis of the turbulence driving mechanisms at the LFS, we assess the role of KH, by considering a two-dimensional simulation where we remove the KH instability drive, i.e., we replace $\phi$ in the $[\phi,\omega]$ term of the vorticity equation (\ref{eq:vort_ph}) by its poloidally averaged counterpart. This simulation (not shown) exhibits an increase of $L_p$ from 18 $\rho_{s0}$ to 30 $\rho_{s0}$, revealing therefore that the KH instability does not drive turbulence, but it plays a role in regulating its saturation level, since it decreases the characteristic gradient lengths in the SOL.

We can therefore conclude that, at the LFS, finite $k_\parallel$ effects decrease the importance of BM and lead to DW driven turbulence whose amplitude is partially regulated by the KH mode at the LFS.
As a comparison, we remark that $k_\parallel$ is set by the ballooning character of the modes in a toroidally limited SOL. This leads to smaller values of $k_\parallel$ and, ultimately, enhances the importance of BM with respect to DW.

We now focus on the HFS, where the DW removal in the nonlinear simulation of \cref{lpfig} significantly decreases $L_p$. This pinpoints the important role of DW as a turbulence drive at this location and rules out BM and KH modes as the main drive of HFS turbulence. The residual turbulence in the DW-suppressed system is driven by the KH mode stabilized by the favourable curvature. This is tested by removing the KH instability drive, and observing that $L_p$ decreases even further to negligible values from approximately $5$ $\rho_{s0}$ to $2$ $\rho_{s0}$.

In addition, two-dimensional simulations (not shown) reveal that $L_p$ increases substantially at the HFS from $L_p \simeq 5$ $\rho_{s0}$ to $L_p \simeq 11$ $\rho_{s0}$ if the curvature term in the vorticity equation is removed, a value in agreement with the estimate in Ref. \cite{Ricci2010}. This shows that favorable curvature has a stabilizing effect on KH. A study on the coupling between the KH instability and BM has been carried out in Ref. \cite{Ricci2006}, where the same effect was noticed.

\begin{figure}
	\centering
	\includegraphics[width=.7\linewidth]{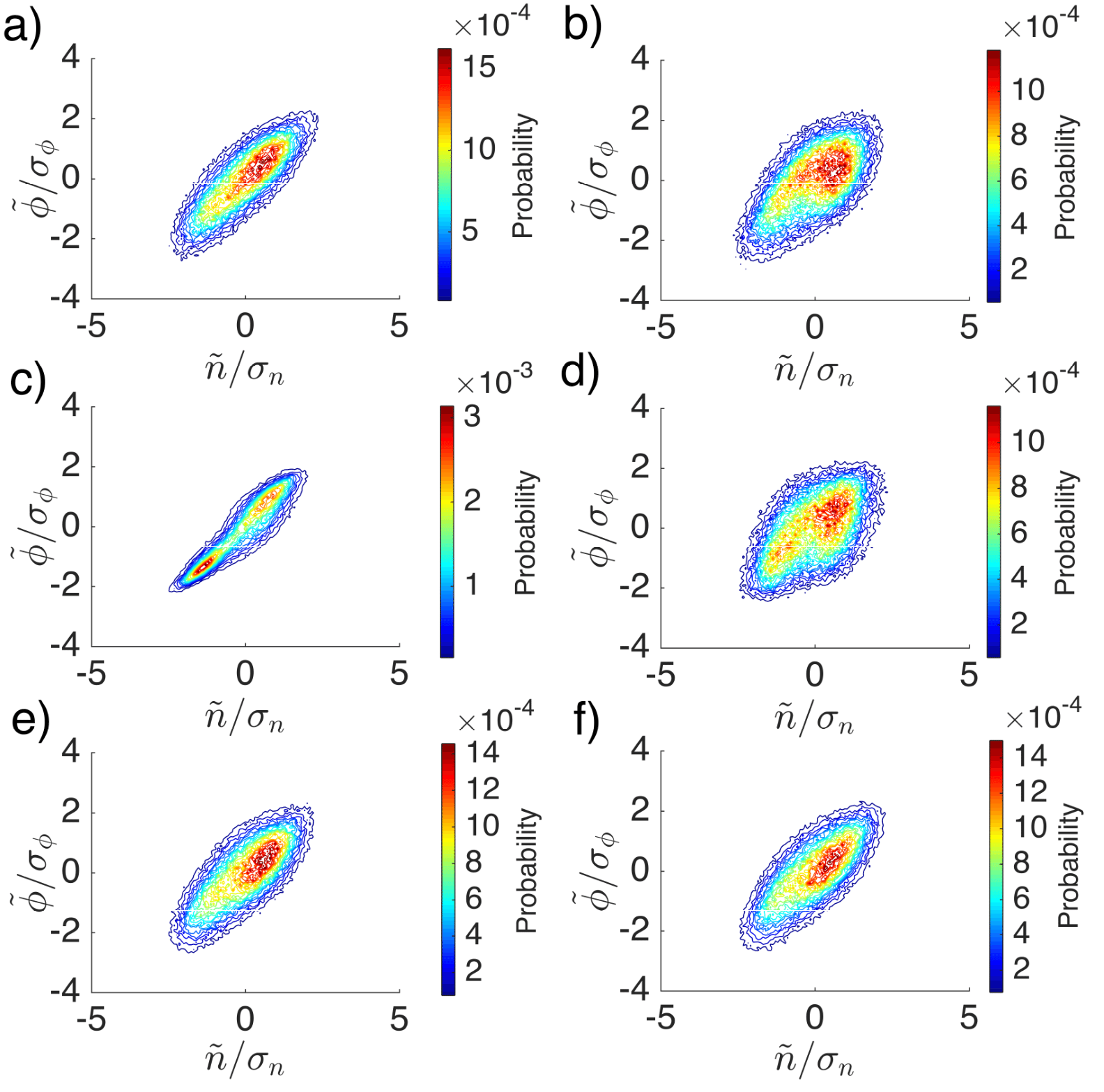}
	\caption{Probability of correlation between density and electric potential fluctuations normalized to their respective standard deviation, resulting from GBS simulations with standard ISTTOK parameters (top), and when DW (middle) or BM (bottom) are removed from the system. The HFS (left) and LFS (right) are shown.}
	\label{crosscoherencefig}
\end{figure}

In order to further justify our conclusions on the turbulent driving mechanisms, we analyse the simulation results by evaluating the cross-coherence and phase-shift between $\tilde n$ and $\tilde \phi$. Here, $\tilde n$ denotes the density fluctuations, defined by $\tilde n = n - \bar n$, with $\bar n$ the time averaged density. An analogous definition is used for the other quantities.
Figure \ref{crosscoherencefig} (top panels) displays the cross-coherence between $\tilde n$ and $\tilde \phi$ for a standard ISTTOK simulation at the radial location $x=x_s+5$ $\rho_{s0}$ and midway toroidally between the two limiter faces at $\varphi = \pi$. The fluctuations are normalized to their standard deviation. Since DW are characterized by an almost adiabatic electron response, a higher correlation between $\tilde \phi$ and $\tilde n$ is expected in DW-driven turbulence with respect to BM-driven turbulence. Indeed, as shown in \cref{crosscoherencefig}, the correlation is strong at the LFS, and even stronger at the HFS, which clearly points to a DW character of turbulence at this location, where the BM  interchange drive is not present.

We also perform a cross-coherence analysis for the three-dimensional simulations where DW, and more specifically the diamagnetic terms $T_e \nabla_\parallel \ln n$ and $1.71 \nabla_{\parallel} T_e$ in Ohm's law (\ref{eq:ohm_ph}), are removed from the system, and for three-dimensional simulations where the BM drive, the curvature term in the vorticity equation (\ref{eq:vort_ph}), is neglected, yielding the middle and bottom panels of \cref{crosscoherencefig} respectively. One observes that BM removal does not affect the correlation at the HFS, and increases it at the LFS (as compared with a standard simulation), as expected from the DW nature of turbulence at the HFS and the mixed BM and DW nature at the LFS. On the other hand, removing DW has the effect of increasing the correlation at the HFS. As a matter of fact, the KH instability that drives transport at the HFS in DW-suppressed turbulent simulations leads to a high correlation between $\tilde n$ and $\tilde \phi$.

We now turn our attention to the phase-shift $-\pi < \delta < \pi$ between $\tilde n$ and $\tilde \phi$, which is expected to be large and close to $\pi/2$ in BM turbulence where, according to \cref{eq:vort_ph}, neglecting $k_\parallel=0$, temperature fluctuations, and KH effects, we have

\begin{equation}
\gamma\nabla_\perp^2 \tilde \phi \sim 2 \omega_{ci}\frac{T_{e}+T_{i}}{e n}C(\tilde n),
\label{eqphaseshift}
\end{equation}

\begin{figure}
	\centering
	\includegraphics[width=.7\linewidth]{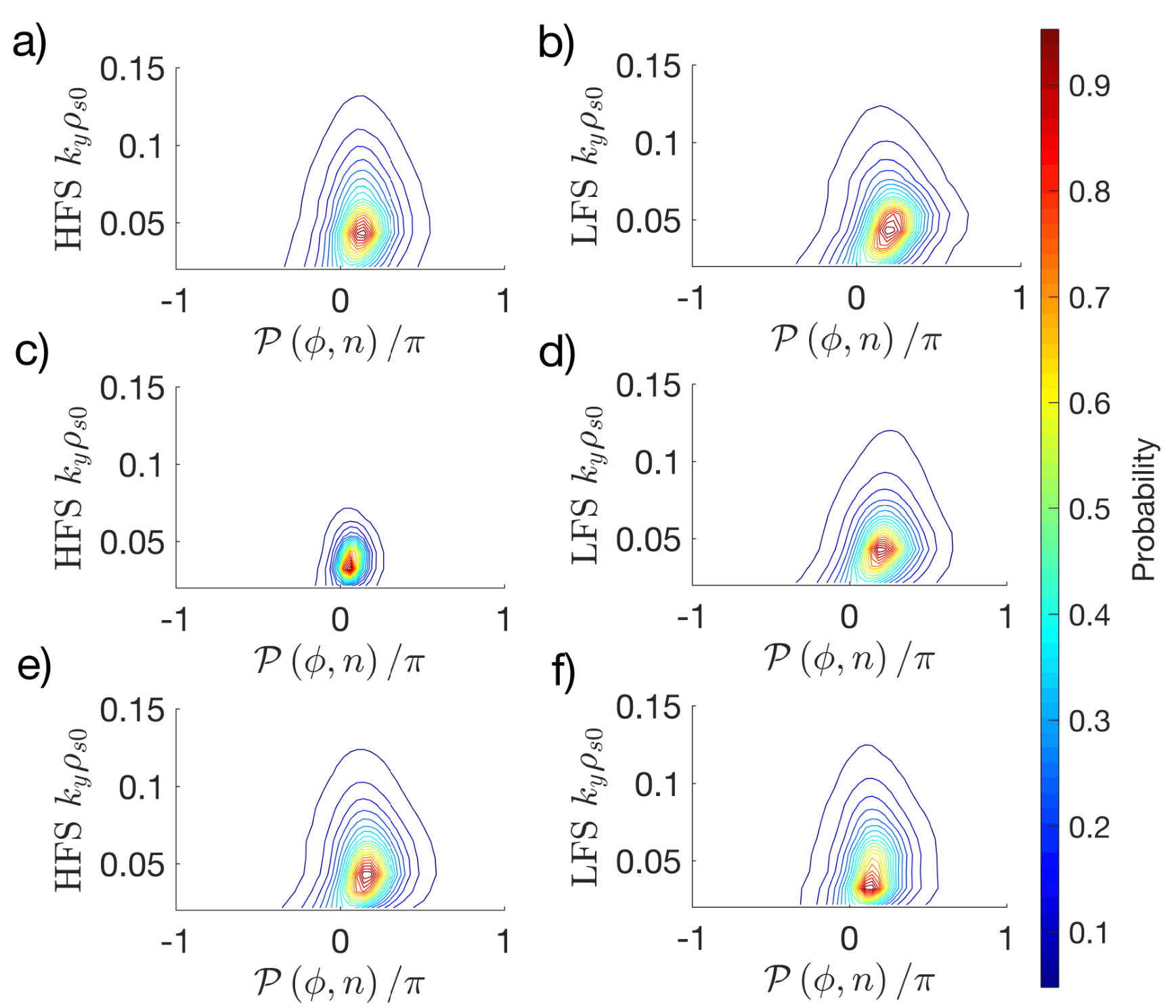}
	\caption{Phase-shift probability between density and electric potential fluctuations, resulting from GBS simulations with standard ISTTOK parameters (top), and when DW (middle) or BM (bottom) are removed from the system. The HFS (left) and LFS (right) are shown.}
	\label{phaseshiftfig}
\end{figure}

\noindent and small in DW driven turbulence, where neglecting electron inertia, temperature fluctuations, and viscous $G_e$ terms we have instead in \cref{eq:ohm_ph}
\begin{equation}
    \nabla_\parallel \tilde \phi \sim \frac{T_{e}}{e n}\nabla_\parallel\tilde n.
\end{equation}

In \cref{phaseshiftfig} we show $\delta$ at $x = x_s+5$ $\rho_{s0}$ and $\varphi=\pi$ (as in \cref{crosscoherencefig}), by performing the Fourier transform of $\tilde \phi$ and $\tilde n$ along $\theta$, on a domain with extension $\Delta \theta = \pi/2$ centered at $\theta=0$ for the LFS, and $\theta = \pm \pi$ for the HFS, and computing the phase shift between these two quantities as a function of $k_y$. The phase shifts evaluated with a frequency of $10^3 c_{s0}/R$, during a time span of $R/c_{s0}$, are then binned as a function of $k_y$ with the proper weight given by the power spectral density of $\tilde \phi$ and $\tilde n$ fluctuations. The results of this test, shown in \cref{phaseshiftfig}, are not particularly clear. In fact, the phase-shift between $\tilde \phi$ and $\tilde n$ is small both at LFS and HFS. Similarly small values are observed if BM and DW drive are removed from the simulation. In fact, \cref{eqphaseshift} is too simplistic to study the phase shift between $\tilde \phi$ and $\tilde n$. The short connection length of our configuration introduces finite $k_\parallel$ effects, that tend to reduce the phase shift. We have highlighted these effects by performing two-dimensional simulations (not shown) with an increasing connection length, and observing that $\delta$ tends to the expected value of $\pi/2$ only when the connection length approaches infinity.

\section{Turbulence Saturation Mechanisms}
\label{sat_section}

Having identified the nature of the linear turbulent drive at different locations, we now turn to the investigation of the mechanisms that saturate the growth of the linearly unstable modes. While a number of saturation mechanisms have been proposed (for a recent review see Ref. \cite{Myra2015a}), it has been shown that the growth of a secondary KH instability and the gradient removal mechanism, i.e., the saturation of the linear mode due to the non-linear local flattening of the driving plasma gradients, are the main saturation mechanisms in the case of DW and BM driven turbulence. Moreover, analytical estimates and numerical simulations suggest that the gradient removal saturation mechanism is present when $\sqrt{k_y L_p} \lesssim 3$ \cite{Ricci2013a}. In our nonlinear simulations, $\sqrt{k_y L_p} \simeq 1$ at the HFS, and $\sqrt{k_y L_p} \simeq 2.2$ at the LFS, points to the gradient removal mechanism as the one at play at the HFS, and partially contributing to the saturation of the unstable modes at the LFS where KH also plays a role in the saturation of the DW, as confirmed in the test described in \cref{sec:identification}. 

When turbulence is saturated by the gradient removal mechanism, the characteristic pressure gradient length $L_p$ in the SOL can be derived by stating that the growth of the linearly unstable modes saturates when the radial gradient of the perturbed pressure becomes comparable to the radial gradient of the background pressure $d\overline p/dx\sim d \tilde p/dx$, which can also be written as

\begin{equation}
    k_x \tilde p \sim \frac{\overline p}{L_p}.
\end{equation}

Following non-local linear theory as outlined in Refs \cite{Ricci2009a,Rogers2005}, for DW and BM respectively, we estimate the radial wavenumber as

\begin{equation}
    k_x\sim \sqrt{\frac{k_y}{L_p}}.
    \label{gr9}
\end{equation}

To estimate the balance between the pressure flux and the parallel losses at the limiter plates, we combine Eqs. (\ref{eq:cont_ph}) and (\ref{eq:te_ph}), and ignore the curvature and diffusion terms, to derive the leading order pressure equation

\be
    \frac{\partial p}{\partial t} = - \frac{c}{B} [\phi, p] - \nabla_\parallel (p V_{\parallel e}).
    \label{eq:press}
\ee

Writing $[\phi,p] = \nabla \cdot \bm \Gamma$, we time average \cref{eq:press}, integrate it along a magnetic field line, and neglect the pressure flux in the poloidal direction $\overline \Gamma_y$ with respect to the turbulent radial flux $\overline \Gamma_x = c\overline{\tilde p {\partial_y \tilde \phi}}/B \sim c k_y \overline{\tilde{\phi} \tilde{p}}/B$.
In addition, estimating the parallel losses at the limiter as $\l.\overline{p V_{\parallel e}}\r|_{\text{limiter}} \simeq \overline p~\overline c_s$, we obtain

\begin{equation}
    \frac{\partial \overline \Gamma_x}{\partial x} \sim - \frac{\overline p~\overline c_s}{2 \pi R}.
    \label{gr4}
\end{equation}

Finally, estimating the electrostatic potential $\tilde \phi$ by neglecting the $k_\parallel$ term in the pressure equation (\ref{eq:press}) as $\tilde \phi\sim B\gamma \tilde p L_p / (R \overline p k_y c)$, and with $\partial_x \overline \Gamma_x \sim \overline \Gamma_x /L_p$, we have

\begin{equation}
    L_p = \frac{R}{c_s}\left(\frac{\gamma}{k_y}\right)_{\text{max}},
    \label{gr10}
\end{equation}

\noindent where $\gamma/k_y$ is maximised over all possible instabilities present in the system. In practice, having fixed $\theta$, the solution of \cref{gr10} requires the evaluation of the linear growth rate $\gamma$  as a function of $k_y$ and $L_p$ from the linear dispersion relation associated with the drift-reduced Braginskii system. We then seek the value of $k_y$ that yields the largest ratio $\gamma/k_y$ for each $L_p$, and we obtain the value of $L_p$ that satisfies \cref{gr10} using Muller's secant method \cite{Muller1956}. A linear code was used to obtain $\gamma$ and $k_y$ for the different unstable modes \cite{Mosetto2012}. Here, a Robin boundary condition \cite{Lanzani2005} is implemented that mimics the dynamics of the different fields at the sheath entrance in the non-linear simulations.

The $L_p$ solution of \cref{gr10} for ISTTOK parameters is shown in \cref{lpfig} as a function of $\theta$ (green dashed line). The agreement with the simulation results is particularly good at the HFS, while at the LFS it overestimates $L_p$ by 25\% (as expected from KH having a role in saturating turbulence).

\begin{figure}
	\centering
	\includegraphics[width=.99\linewidth]{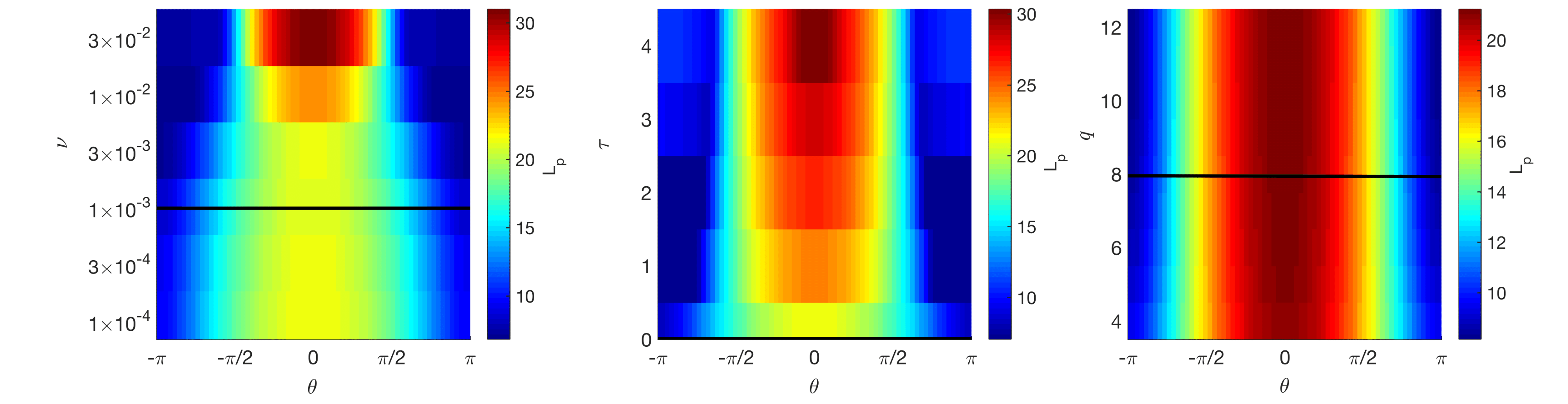}
	\caption{Equilibrium pressure scale-length, $L_p$, as a function of the poloidal angle $\theta$ and normalised resistivity, $\nu$ (left panel), ion to electron temperature ratio, $\tau$ (middle panel), and safety factor, $q$ (right panel). The black lines represent the standard ISTTOK case.}
	\label{linearlp}
\end{figure}

Using the result of \cref{gr10}, we also estimate $L_p$ as a function of the resistivity $\nu$, ion to electron temperature ratio $\tau$, and safety factor $q$ in order to assess the dependence of the SOL radial pressure profile on these parameters. The results of this estimate are shown in \cref{linearlp}, and reveal that $L_p$ depends weakly on the safety factor $q$, while it increases for increasing values of $\nu$ and $\tau$.

Equation (\ref{gr10}) allows us to further confirm the ISTTOK turbulent regimes identified in \cref{sec:identification}, and extend this analysis to a wide parameter space. In fact, having estimated $L_p$ as a function of $\tau, \nu$, and $q$, one can evaluate the growth rate of the Resistive BM, Inertial BM, Resistive DW, and Inertial DW instabilities. We note that the resistive branch of BM and DW is due to the presence of resistivity ($\nu$) in Ohm's law, \cref{eq:ohm_ph}, while an inertial branch of BM and DW is made unstable by electron inertia ($m_e$) effects. Therefore, the growth rate of the resistive BM and DW can be found by neglecting $m_e$ in \cref{eq:ohm_ph}, while the inertial instability is evaluated by neglecting $\nu_e$ in \cref{eq:ohm_ph}.
In order to identify the turbulent regimes we evaluate the growth rate of the four instabilities above at the $k_y$ and $L_p$ that solve \cref{gr10}. Turbulence is expected to be driven by the instability that has the largest linear growth rate.

\begin{figure}
	\centering
	\includegraphics[width=.99\linewidth]{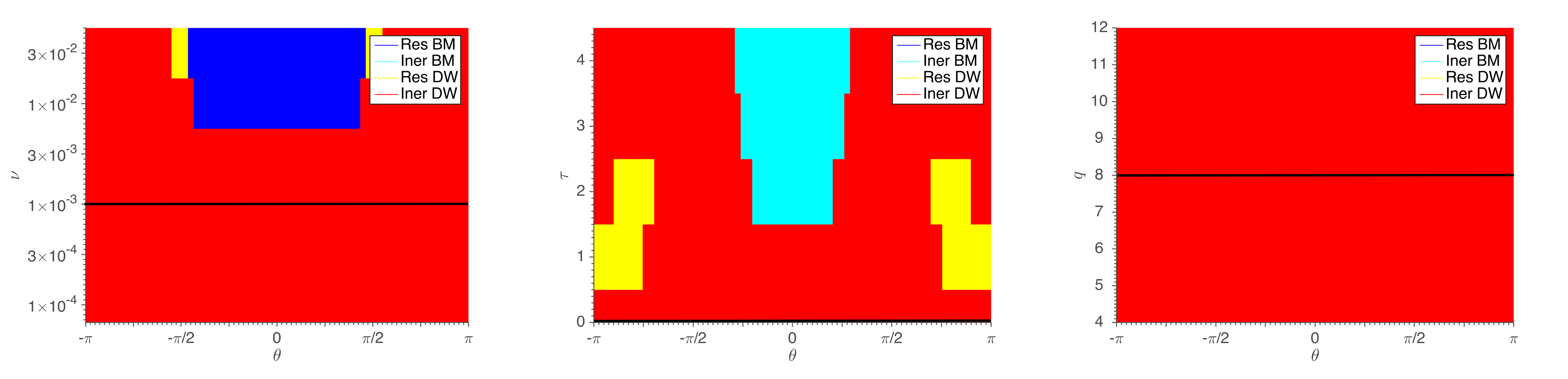}
	\caption{Turbulent regimes as a function of the poloidal angle $\theta$ and normalised resistivity, $\nu$ (upper left panel), ion to electron temperature ratio $\tau$ (middle panel), and safety factor $q$ (right panel). The Resistive BM driven turbulence is in dark blue, Inertial BM in light blue, Resistive DW in yellow, and Inertial DW in red. The black lines represent the standard ISTTOK case.}
	\label{turbfig}
\end{figure}

The turbulence regimes are shown in \cref{turbfig}, where Inertial DW drives turbulence at all poloidal angles for typical ISTTOK parameters. An increase of the resistivity $\nu$ from the typical ISTTOK standard simulation value, $\nu \sim 1 \times 10^{-3}$, to $1 \times 10^{-2}$ leads to the Resistive BM at the LFS, while for $\tau > 1$ a transition to the Inertial BM is also seen near $\theta = 0$. For the case of $1 < \tau < 2$, Resistive DW drive turbulence at the HFS, and the turbulent regime is not affected by the safety factor in a wide range of values $(4 < q < 12)$.

\section{Comparison with Experimental Results}
\label{exp_comparison}

To compare our numerical results with experimental measurements we consider an ISTTOK discharge with density $n=4 \times 10^{18}$ $\text{m}^{-3}$ and $q=10$ at the LCFS. 
The experimental measurements were obtained with a multi-pin Langmuir probe measuring simultaneously the floating potential $V_f$ and ion saturation current $I_{sat}$. The probe was moved from shot-to-shot along the radial direction and measurements were taken at $r-a=0,~5$ and $10$ mm \cite{Silva2015}.
We note that experimental measurements show the presence of a shear layer inside the last-closed flux surface, $-10 < r-a < 0$ mm in ISTTOK. While the statistical properties of the fluctuations are locally affected by the shear layer \cite{Sanchez2000}, our measurements are not influenced by its presence since they are taken at $r-a \ge 0$.
The experimental uncertainty was estimated by performing three different discharges with the same parameters. From the simulation results we evaluate $I_{sat}=e n c_s A$ ($A$ being the probe area) and $V_{f}= \phi- \Lambda T_e$.

\begin{figure}
	\centering
    \includegraphics[width=.99\linewidth]{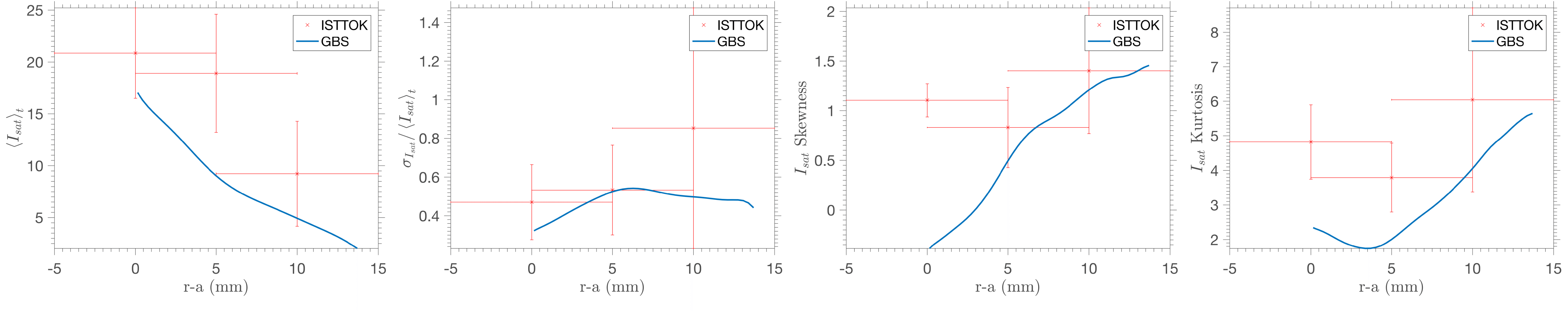}
	\caption{Statistical moments of $I_{sat}$ from the experiment (red) and simulation (blue). From left to right: mean, standard deviation, skewness, and kurtosis.}
	\label{statIsat}
\end{figure}

First, we focus on the $I_{sat}$ statistical moments in \cref{statIsat}.
The temporal mean of $I_{sat}$ is monotonically decreasing for increasing radial locations, both in the simulations and in the experiments. However, the large uncertainty does not allow us to compare reliably the $I_{sat}$ gradient scale length. The standard deviation shows that fluctuations are large, approximately 50\%, throughout the SOL both in the experiment and simulation, as it is typically observed in the SOL of fusion devices. The simulation results show a monotonically increasing skewness, as expected from previous SOL studies \cite{Nanobashvili2007,Xu2005,Xu2010,Sanchez2000}. On the other hand, in ISTTOK, we find a rather large value ($\simeq 1$) of the skewness at the LCFS. The skewness (as well as the kurtosis) shows a better agreement between simulations and experiment in the far SOL.

\begin{figure}
	\centering
    \includegraphics[width=.9\linewidth]{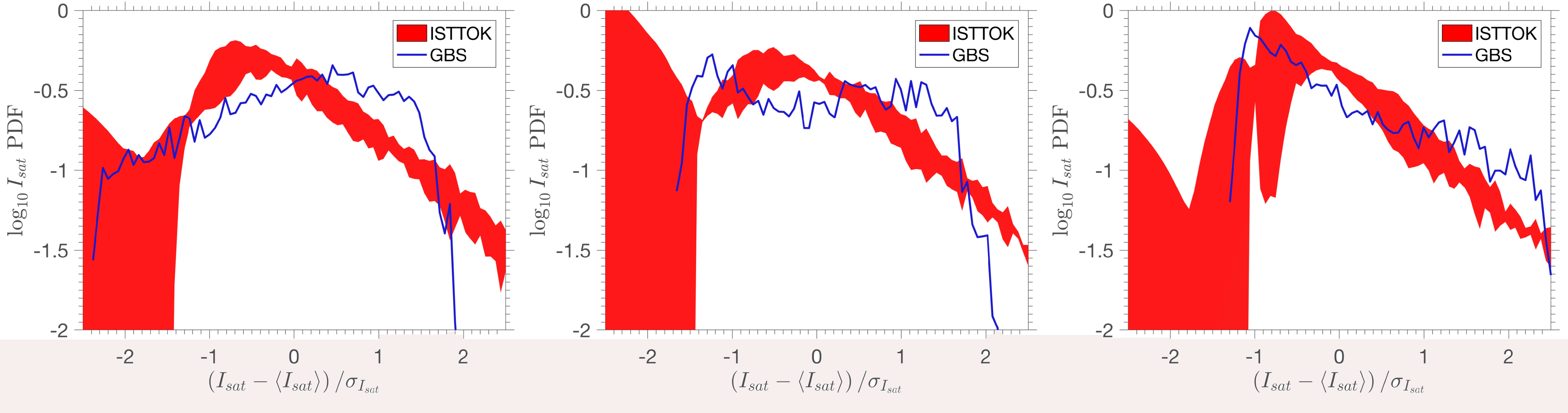}
	\caption{$I_{sat}$ PDF from the experiment (red) and simulation (blue), at $r-a=0$ mm (left), $r-a=5$ mm (center), and $r-a=10$ mm (right).}
	\label{pdffig1}
\end{figure}

These observations are confirmed by the comparison of the $I_{sat}$ probability distribution function (PDF) shown in \cref{pdffig1}. In all cases the $I_{sat}$ PDFs deviate strongly from a Gaussian distribution and we observe that the $I_{sat}$ PDF is considerably more skewed in the experiment than in the simulation at the LCFS. The level of agreement increases while moving towards the far SOL. The discrepancy between simulation and experimental results in the proximity of the LCFS might be due to intermittent events occurring in ISTTOK inside the LCFS. These events are not captured by the simulation that cannot properly describe the coupling with core physics.

\begin{figure}
	\centering
    \includegraphics[width=.9\linewidth]{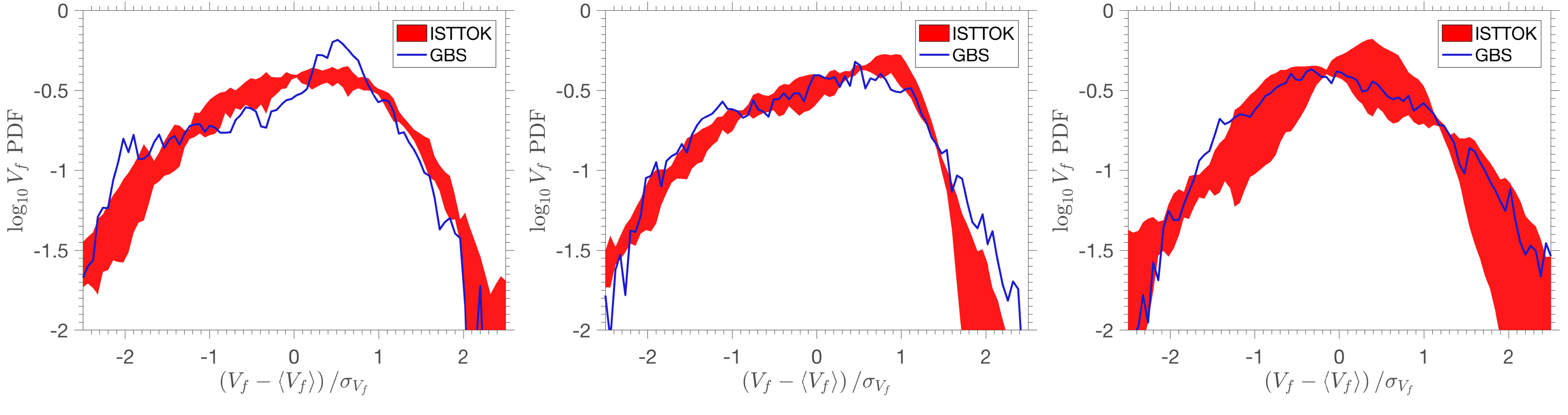}
	\caption{$V_f$ PDF from the experiment (red) and simulation (blue), at $r-a=0$ mm (left), $r-a=5$ mm (center), and $r-a=10$ mm (right).}
	\label{pdffig2}
\end{figure}

As opposed to $I_{sat}$, the $V_f$ PDFs show agreement with the simulation results within the error bars for the different radial locations (see \cref{pdffig2}). We remark that the $V_f$ PDFs are rather symmetric, possibly due to the bipolar nature of $V_f$ associated with the intermittent events, and display Gaussian properties \cite{Furno2011}.

\begin{figure}
	\centering
	\includegraphics[width=.99\linewidth]{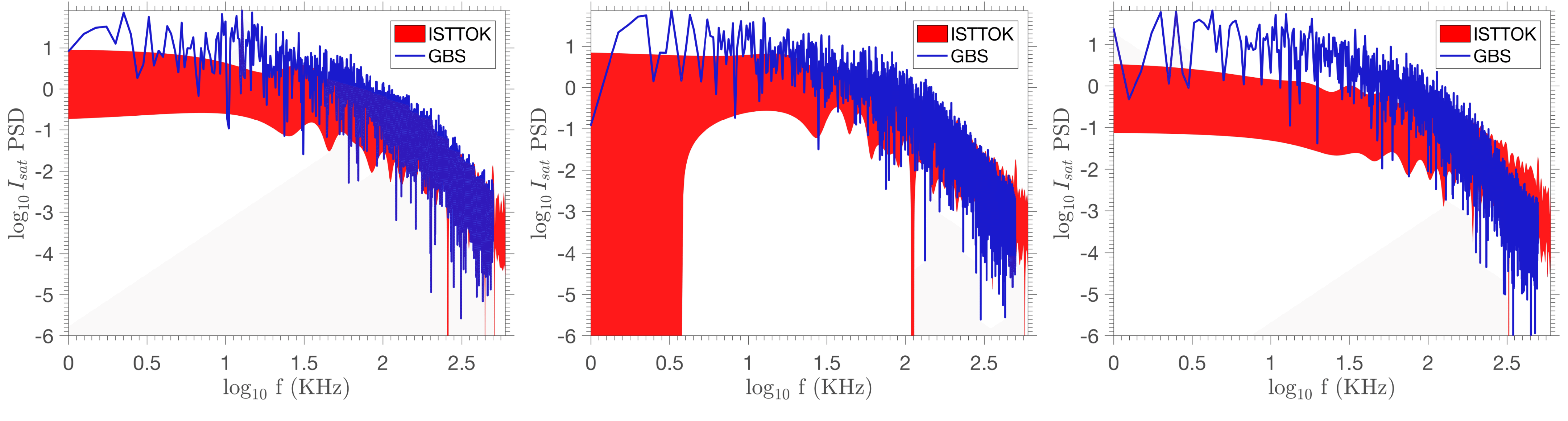}
	\caption{$I_{sat}$ power spectra from the experiment (red) and simulations (blue), at $r-a=0$ mm (left), $r-a=5$ mm (center), and $r-a=10$ mm (right).}
	\label{powerlawfig1}
\end{figure}

\begin{figure}
	\centering
	\includegraphics[width=.99\linewidth]{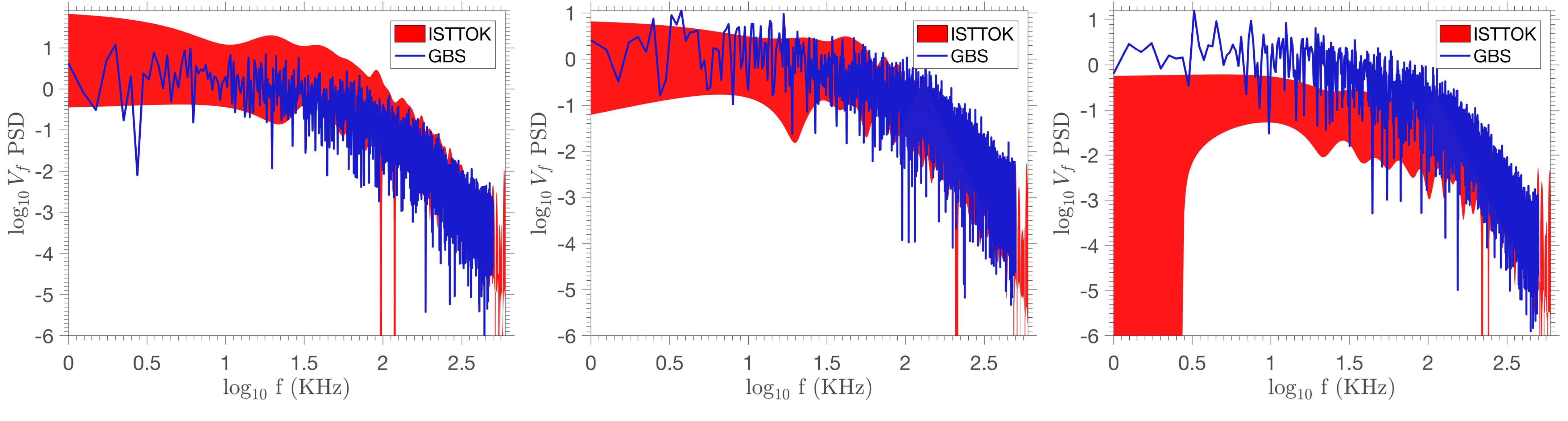}
	\caption{$V_f$ power spectra from the experiment (red) and simulations (blue), at $r-a=0$ mm (left), $r-a=5$ mm (center), and $r-a=10$ mm (right).}
	\label{powerlawfig2}
\end{figure}

We then consider the  $I_{sat}$ and $V_f$ power spectral density, evaluated as the square of the absolute value of the temporal Fourier transform. These are shown in \cref{powerlawfig1,powerlawfig2} for $I_{sat}$ and $V_f$ respectively. In all cases, the power spectra are approximately flat for frequencies $\lesssim 20$ kHz, a typical behavior observed in tokamak SOL turbulence \cite{Carreras1999}. 
At higher frequencies, we compare the spectrum decay index between ISTTOK and GBS profiles. Focusing on the region $50 < f < 300$ kHz, we assume a power-law of the form $A f^\mu$ with $A$ a constant, $f$ the frequency, and $\mu$ the decay index. We find that the $I_{sat}$ power spectra show a sharper decrease in the simulation, as compared to the experiment, while for $V_f$, we find a sharper decrease in the experimental values. Quantitatively, at $r-a =5$ mm, the experiment and simulation $I_{sat}$ spectral index are $\mu_{\text{exp}}=-1.69 \pm 0.32$ and $\mu_{\text{sim}}=-2.20\pm0.03$ respectively, while for $V_f$ we find $\mu_{\text{exp}}=-2.07\pm0.34$ and $\mu_{\text{sim}}=-1.64\pm0.03$.

Finally, in order to compare the experimental pressure gradient lengths $L_p$ with the ones discussed in \cref{sat_section}, experimental measurements of $n$ and $T_e$ were taken using sweeping Langmuir probes with 3 mm radial resolution. Experimental measurements suggest that $L_p$ is independent of $q$ for a wide range of values ($L_p = $ 4.8, 4.5, 4.3 $\rho_{s0}$ for $q=$ 7, 10, 13 respectively), a behavior in agreement with simulation results (see \cref{linearlp}).
However, the experimental value of $L_p \simeq 4.5$ $\rho_{s0}$ at the LFS differs from the one predicted in simulation results by a factor larger than three. In fact, in the ISTTOK standard simulation we have $L_p\simeq 15$ $\rho_{s0}$ at the LFS (see \cref{lpfig}). This might be due to the presence of the outer wall in the experiment that acts effectively as a plasma sink and reduces $L_p$. Its presence is not accounted for in the GBS simulations, which considers a large radial domain extension.

We remark that the longer pressure scale length observed in the experiment strengthens our theoretical observation that KH is not the driving mechanism of turbulence in ISTTOK SOL. A straightforward comparison of the linear growth rate of the KH instability, $\gamma_{\text{KH}} \sim 0.025 (c/B) \phi/L_p^2$, of the DW, $\gamma_{\text{DW}} \sim 0.25 c_s/L_p$, and of the BM, $\gamma_{\text{BM}} \sim c_s/\sqrt{R L_p}$, (see Ref. \cite{Ricci2010}), shows that 
 $\gamma_{\text{KH}}/\gamma_{\text{DW}}\sim\gamma_{\text{KH}}/\gamma_{\text{BM}}\sim 10^{-2}$.

\section{Conclusions}
\label{sec:conclusion}

The present paper addresses the study of plasma turbulence in a poloidally limited SOL, using linear calculations and non-linear simulations based on the drift-reduced Braginskii equations. We focus our investigations on the parameters of the ISTTOK tokamak and compare our theoretical results with experiments carried out there.

Significant differences are found with respect to a toroidally limited SOL. Because of the presence of the poloidal limiter that avoids the connection between the LFS and HFS, a clear poloidal asymmetry is observed, with the time-averaged pressure scale length considerably shorter at the LFS compared with the HFS. Due to the short connection length and related steep pressure gradients, the role of DW is enhanced with respect to the toroidally limited case. In fact, for the typical ISTTOK parameters, we identify DW as the main linear instability drive both at the LFS and HFS, where we also find KH to play a non-negligible role in saturating turbulence.

The pressure scale length obtained from the non-linear simulations shows a remarkable agreement with estimates based on the saturation of the unstable linear modes due to the non-linear local flattening of the driving plasma gradients at the HFS. The agreement decreases at the LFS due to the aforementioned role of the KH instability in setting the turbulence amplitude.

The comparison of the statistical properties of turbulence shows a good agreement between experimental and numerical results particularly in the far SOL. Intermittent events observed in ISTTOK in the near SOL are not captured by the simulation. On the other hand, possibly because of the interaction of the plasma with the wall, the characteristic pressure scale gradient length found in the simulation is considerably larger than that measured in the experiment.

\section{Acknowledgments}

Part of the simulations presented herein were carried out using the HELIOS supercomputer system at Computational Simulation Centre of International Fusion Energy Research Centre (IFERC-CSC), Aomori, Japan, under the Broader Approach collaboration between Euratom and Japan, implemented by Fusion for Energy and JAEA; and part were carried out at the Swiss National Supercomputing Centre (CSCS) under Project ID s549. This work has been carried out within the framework of the EUROfusion Consortium and has received funding from the Euratom research and training programme 2014-2018 under grant agreement No 633053, and from Portuguese FCT - Fundação para a Ciência e a Tecnologia, under grant PD/BD/105979/2014. The views and opinions expressed herein do not necessarily reflect those of the European Commission.

\bibliographystyle{unsrt}
\bibliography{library}

\end{document}